\documentclass[12pt,epsfig]{article}

\usepackage{amssymb,epsfig}

\newcommand{\AmS}{{\protect\the\textfont2
  A\kern-.1667em\lower.5ex\hbox{M}\kern-.125emS}}

\begin{document}

\begin{titlepage}

\begin{flushright}
\begin{tabular}{l}
  hep-ph/0210nnn
\end{tabular}
\end{flushright}
\vspace{1.5cm}

\begin{center}

{\LARGE \bf
Charge and Spin Asymmetries from Pomeron-Odderon Interference}

\vspace{1cm}

{\sc P. H\"agler}${}^{1}$,
{\sc B.~Pire}${}^{2}$,
{\sc L.~Szymanowski}${}^{2,3}$ and
{\sc O.V.~Teryaev}${}^{2,4}$
\\[0.5cm]
\vspace*{0.1cm} ${}^1${\it
   Institut f{\"u}r Theoretische Physik, Universit{\"a}t
   Regensburg, \\ D-93040 Regensburg, Germany
                       } \\[0.2cm]
\vspace*{0.1cm} ${}^2$ {\it
CPhT, {\'E}cole Polytechnique, F-91128 Palaiseau, France\footnote{
  Unit{\'e} mixte C7644 du CNRS.}
                       } \\[0.2cm]
\vspace*{0.1cm} ${}^3$ {\it
 So{\l}tan Institute for Nuclear Studies,
Ho\.za 69,\\ 00-681 Warsaw, Poland
                       } \\[0.2cm]
\vspace*{0.1cm} ${}^4$ {\it
Bogoliubov Lab. of Theoretical Physics, JINR, 141980 Dubna, Russia
                       } \\[1.0cm]

\vskip2cm
\end{center}
We study Pomeron-Odderon interference effects giving rise to charge and
single-spin
asymmetries in diffractive electroproduction of a
 $\pi^+\;\pi^-$ pair. We calculate these asymmetries
in the Born approximation of QCD in a kinematical domain accessible to HERA
experiments.


\vspace*{1cm}

\end{titlepage}

\section{Introduction}

Hadronic reactions at low momentum transfer and high energies (for
charge-even exchange)
are described in QCD  in terms of
QCD-Pomeron described by the BFKL equation \cite{BFKL}.
The charge-odd exchange is less well understood although the corresponding
BKP equations \cite{BKP} have attracted much attention
recently \cite{Levodd,JW,Vacca1,Korch}, thus
reviving the relevance of phenomenological studies of the Odderon exchange
pointed out years ago in Ref. \cite{LN}.
Unfortunately, the recent studies of specific channels where the
QCD Odderon contribution is expected
to be singled out have turned out to be very disappointing.
In particular, recent experimental studies at HERA of
exclusive $\pi^0$ photoproduction  \cite{Olsson}  indicate a very
small cross section for this process which stays in contradiction with
theoretical predictions based on the stochastic vacuum model \cite{Dosh}.

The general feature
of all
meson production processes is that scattering amplitude describing Odderon
exchange enters quadratically in the cross section.
This observation lead to the suggestion
in Ref. \cite{Brodsky}, that the study of observables where Odderon effects
are present  at the amplitude level - and not at the squared amplitude
level - is
mandatory to get a convenient sensitivity to a rather small normalization
of this contribution.
 This may be achieved by means of  charge asymmetries, as for
instance in open charm production \cite{Brodsky}. Since the final state
quark-antiquark pair has no definite charge parity both Pomeron and Odderon
exchanges
contribute to this process. Another example \cite{Nikolaev} is the charge
asymmetry in soft
photoproduction of two pions.
Bearing in mind perturbative QCD (pQCD) description, we calculated
the "hard" analogs
of these asymmetries \cite{HPST}, and supplemented them by the
single spin asymmetries,
which may be studied at HERA with polarized lepton beam \cite{HPST1}.

\section{Method and Results}
We consider the process
$e^-(p_e) N(p_N) \to e^-(p_e^{\prime})\pi^+(p_+) \pi^-(p_-)
N^{\prime}(p_N^{\prime})$. The application of pQCD for the calculation of a
part of this process is justified by the presence of a
 hard scale: the squared mass $-Q^2=-(p_e - p_e^{\prime})^2$
 of the virtual photon,
$Q^2$ being of the order of a few GeV$^2$.
The amplitude of this process (Fig.1)
\begin{figure}[htb]
\vspace{6pt}
\centerline{\epsfxsize8.0cm\epsffile{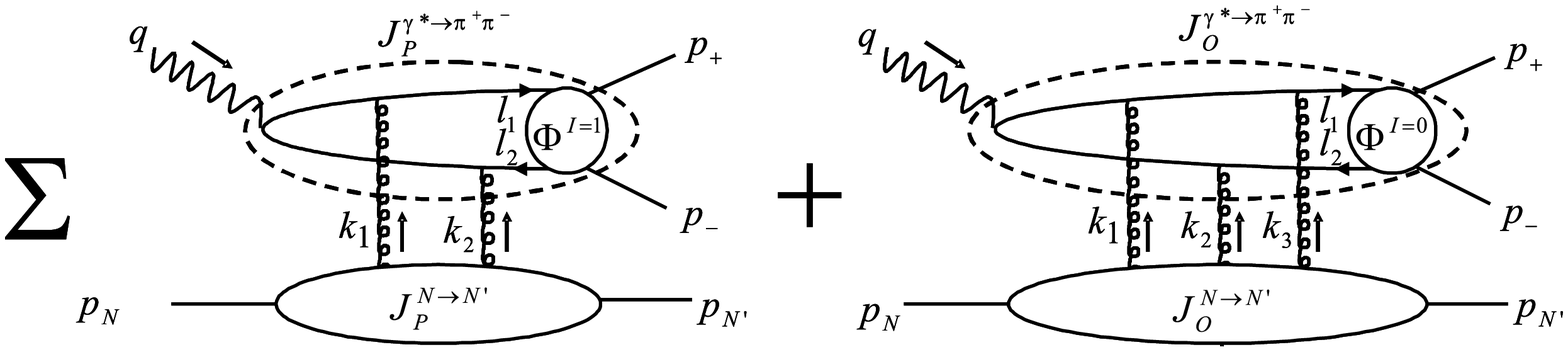}}
\caption{Pomeron and Odderon amplitudes within QCD factorization}
\label{graphs}
\end{figure}
includes the convolution of a
perturbatively calculable hard subprocess with two non-perturbative inputs,
the 2-pion generalized distribution
amplitude (GDA) and the Pomeron-Odderon (P/O)
proton impact factors. Since
the $\pi^+\pi^-$ system is not a  charge parity
eigenstate, the GDA includes two charge parity components and allows for
a study of the corresponding interference term. The relevant GDA is
just given by the light cone wave function of the two pion system \cite{DGPT}.

\subsection{Charge asymmetry}
We define the  forward-backward or charge asymmetry
$A(Q^{2},t,m_{2\pi }^{2},y,\alpha )$ by
\begin{eqnarray}
\label{ca}
 \frac{\sum\limits_{\lambda =+,-}\int \cos
\theta \,d\sigma (s,Q^{2},t,m_{2\pi }^{2},y,\alpha ,\theta ,\lambda )}{%
\sum\limits_{\lambda =+,-}\int d\sigma (s,Q^{2},t,m_{2\pi }^{2},y,\alpha
,\theta ,\lambda )}
\end{eqnarray}
as a weighted integral over polar angle $\theta$ of the relative momentum of
two pions. Although this asymmetry depends on the full set of the kinematical
variables, different dependencies, due to factorization, come from different
sources.
The most clean one are the dependencies on $Q^2$, coming
from the hard subprocess, and on dipion mass $m_{2\pi}=\sqrt{(p_+ + p_-)^2}$,
coming from GDA. The specific form of
$m_{2\pi}$ dependence is explained by the fact, that the phase of
GDA\cite{DGPT} should add to the phase shift between Pomeron and Odderon.
The $Q^2$ dependence, which is a subject to corrections from
BFKL, BKP and ERBL evolution, is due to the different $Q^2$ dependence of
Pomeron and Odderon coefficient functions (perturbative impactfactors),
the latter having the extra propagators, leading to the decrease of
asymmetry with $Q^2$. The typical scale of $\alpha_s$ is determined by
the gluons transverse momenta $\vec{k_i}$ (see Fig.1)
and we checked that the result is not
changed substantially if "hard" $\alpha_S(Q^2)$ is used for coefficient
function and "soft" $\alpha_S \sim 0.5$ is used for non-perturbative
proton impactfactors.
\begin{figure}[htb]
\vspace{6pt}
\centerline{\epsfxsize8.0cm\epsffile{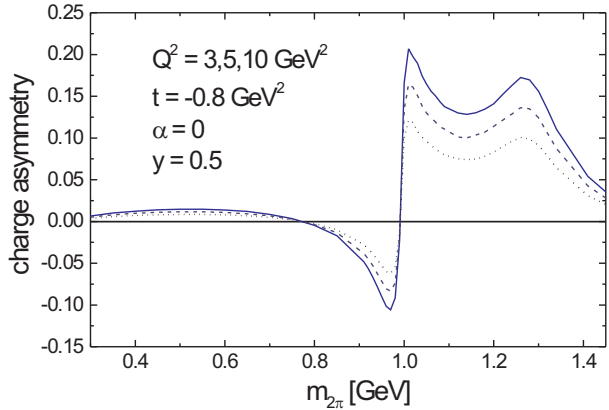}}
\caption{Dependence of charge asymmetry on $m_{2\pi}$ and $Q^2$}
\label{m2pi}
\end{figure}
The latter objects, together with the coefficient functions,
define the dependence of asymmetry
on $t=(p_N-p_N^{\prime})^2$. As to the dependence on the standard
leptonic variables $y$ (the lepton
energy fraction carried by the virtual photon) and $\alpha$
(the angle between lepton and photon scattering planes),
it is determined by the
relative size of the virtual photon helicity amplitudes and allows to estimate
the promising kinematical region \cite{HPST1}.

\subsection{Spin Asymmetry}

The single spin asymmetry  $A_S(Q^{2},t,m_{2\pi }^{2},y,\alpha )$
is defined by
\begin{eqnarray}
\label{sa}
 \frac{\sum\limits_{\lambda
=+,-}\lambda \int \cos \theta \,d\sigma (s,Q^{2},t,m_{2\pi }^{2},y,\alpha
,\theta ,\lambda )}{\ \sum\limits_{\lambda =+,-}\int d\sigma
(s,Q^{2},t,m_{2\pi }^{2},y,\alpha ,\theta ,\lambda )}
\end{eqnarray}
and requires to fix the lepton beam polarization $\lambda$.
Contrary to charge asymmetry, this effect is proportional to
the imaginary, rather than to real part of the interference
term. As the Pomeron amplitude is imaginary
and the Odderon one is real
the relative phase between them is the maximal one for the emergence of
single spin asymmetries \cite{OT01}. The effect should be therefore
maximal for  zero relative phase between isoscalar and isovector
GDA's, providing
a complementary probe. This complementarity can be seen
from the dependence of spin asymmetry on $m_{2\pi}$ and $Q^2$(Fig.2)
\begin{figure}[htb]
\vspace{9pt}
\centerline{\epsfxsize8.0cm\epsffile{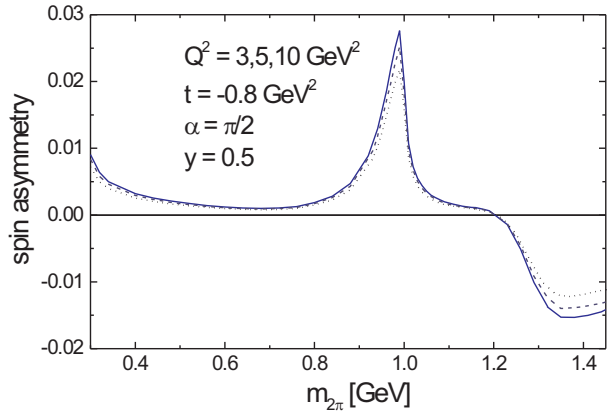}}
\caption{Dependence of spin asymmetry on $m_{2\pi}$ and $Q^2$     }
\label{spm2pi}
\end{figure}
The smaller numerical value of spin asymmetry is to a large extent
due to the kinematical factor $\sqrt{t}$.
This asymmetry may be therefore important in the region of large $t$,
which would become a relevant
QCD scale and allow to have smaller $Q^2$.

\section{Conclusions}

We found that a sizable charge asymmetry may be a useful tool
to look for QCD Odderon contribution\ at HERA.  The spin asymmetry is smaller,
but it can be important at larger $t$ and smaller $Q^2$.
Note finally, that the numerical value of our predictions depend on the adopted
model for proton impact factors, so the observation of the predicted effects
with the different magnitude and/or $t-$dependence might be considered
as an indirect experimental determination of these important non-perturbative
objects.

\end{document}